\documentclass[journal]{IEEEtran}
\usepackage{cite}
\usepackage{amsmath,amssymb,amsfonts}
\usepackage{algorithmic}
\usepackage{graphicx}
\usepackage{booktabs}
\usepackage{array}
\usepackage{algorithm}
\usepackage{hyperref}
\hypersetup{hidelinks=true}
\usepackage{textcomp}

\begin{document}
\newcommand{\manuscriptabstract}{We previously introduced ultrasound autofocusing, an iterative model-based aberration correction technique that estimates local sound speed and incorporates it into beamforming to correct image distortion from heterogeneous media. In this work, we extend ultrasound autofocusing to curvilinear arrays and introduce advancements to the underlying model. A differentiable bent-ray tracing approach accounts for refraction through the transducer lens, while a new adaptive-grid initialization accounts for changes in speckle position with sound speed. The method is validated in silico and in calibrated sound speed phantoms. Our distributed aberration-correction method is then applied to a first large-scale in vivo evaluation comprising 313 liver acquisitions from 76 high-BMI human subjects. In images containing anechoic regions, contrast and CNR improved by 1.38$\pm$1.60 dB (+18.0\%) and 0.09$\pm$0.14 (+10.2\%), respectively. Improvements were also observed in speckle brightness (+20.3\%), coherence factor (+13.1\%), lag-one coherence (+2.7\%), common-midpoint correlation coefficient (+0.7\%), and common-midpoint phase error (-9.1\%; lower is better), with all metric improvements statistically significant. Target structure and visibility also improved significantly. These results demonstrate the potential of ultrasound autofocusing for clinically applicable distributed aberration correction.}
\newcommand{\manuscriptkeywords}{ultrasound autofocusing, differentiable beamforming, aberration correction, acoustic imaging}

\title{Lens-Aware Differentiable Beamforming for\\In Vivo Distributed Aberration Correction with Curvilinear Transducers}

\author{Benjamin N. Frey, Robin van Velzen, Hoda S. Hashemi, Samuel Beuret, Martin Schneider, Alice C. Fan, Christian R. Hoerner, Aya Kamaya, Sergio J. Sanabria, and Jeremy J. Dahl
\thanks{This research was supported in part by NIBIB Grant R01-EB027100.\\
Benjamin N. Frey is with the Dept. of Applied Physics, Stanford University, Stanford, CA, USA.\\
Robin van Velzen is with the Dept. of Biomedical Engineering, Eindhoven University of Technology, Eindhoven, Netherlands.\\
Hoda S. Hashemi, Samuel Beuret, Martin Schneider, Alice C. Fan, Christian R. Hoerner, Aya Kamaya, Sergio J. Sanabria, and Jeremy J. Dahl are with the Dept. of Radiology, Stanford University, Stanford, CA, USA.\\
B. N. Frey (e-mail: benfrey@stanford.edu) and J. J. Dahl (e-mail: jjdahl@stanford.edu) are the corresponding authors.}}

\maketitle

\begin{abstract}
\manuscriptabstract
\end{abstract}

\begin{IEEEkeywords}
\manuscriptkeywords
\end{IEEEkeywords}

\section{Introduction}
\label{sec:introduction}

 Overweight{} and obesity affect nearly 73$\%$ of U.S. adults (43$\%$ worldwide), and this condition is associated with increased adipose tissue~\cite{fryar2026prevalence}. Clinical ultrasound scanners assume a homogeneous tissue sound speed to beamform images, but this assumption can frequently break down in this population due to increased fat layers with varying sound speeds. These variations cause ultrasonic wavefront distortion which leads to B-mode defocusing, reduced lateral resolution, and a decrease in contrast. Distributed aberration correction that can dynamically adjust image reconstruction sound speed locally can provide more effective imaging sessions and downstream pathological decision making.

In abdominal imaging, curvilinear ultrasound arrays are commonly used because they provide the advantage of imaging a wider field of view (FOV) than linear arrays. They also typically employ lower central frequencies to enable deeper image acquisitions. In the abdomen, wavefront aberration arises from path-integrated travel-time deviations, typically induced by the large changes in the speeds of sound of the subcutaneous tissue layers. Furthermore, incorrect beamforming speed of sound can accumulate wavefront distortion over depth, resulting in more apparent degradation in larger and deeper imaging FOVs. Additional significant refraction occurs through the mechanical transducer lens, causing further errors in beamforming. While commercial ultrasound systems commonly account for lens refraction in their beamforming, these calculations typically assume that the refraction through the lens is coupled into a 1540 m/s sound speed medium, resulting in image reconstruction that is still distorted. The technical challenges facing the practical implementation of clinical aberration correction are twofold: (i) errors in the sound speed assumption made by the scanner system result in distorted image reconstruction, and (ii) uncorrected lens refraction challenges accurate sound speed estimation.

There is a wide history of aberration modeling and correction techniques in the literature~\cite{ali2023aberration}. Early approaches to aberration correction involved the estimation of time-delays and amplitude distortions from the wavefronts using near-field phase screen~\cite{o1988phase} and distributed aberration models~\cite{Hinkelman1998abdominal,Mast1998simulation}. More recently, aberration correction based on estimates of the medium's sound speed have shown greater promise than earlier models in correcting image distortion~\cite{jaeger2015full,rau2019ultrasound,ali2022distributed,vraalstad2026coherence}. In these sound speed-based models, propagation delays used for image reconstruction are derived from either a global estimation of sound speed or a spatially varying sound speed field. Jaeger et al. \cite{jaeger2015full} estimated a distributed sound speed map from transmit steering-dependent echo phase shifts and used the resulting map to derive propagation delays for aberration correction. Rau et al.~\cite{rau2019ultrasound} similarly used plane-wave acquisitions and a two-dimensional sound speed map for distributed aberration correction, estimating the map from differential time-of-flight measurements inferred from apparent displacements between images acquired at different transmit angles~\cite{sanabria2018spatial}. Ali et al.~\cite{ali2022distributed} subsequently used pulse-echo tomographic sound-speed estimates in two distributed aberration-correction techniques: delay-and-sum beamforming based on time delays calculated using the eikonal equation and wavefield correlation imaging.

More recently, Simson et al.~\cite{simson2025ultrasound} presented a straight-ray autofocusing model based on automatic differentiation to enable simultaneous sound speed estimation and aberration correction for linear arrays while using common-midpoint phase error as a focusing criterion. The main limitation of this model is that significant refraction occurs through the transducer lens, causing errors in sound speed approximation that were previously not accounted for. Additionally, the grid spacing used for image quality assessment is held constant despite sweeping the global sound speed used for model initialization, creating an initialization bias. In this work, we introduce a number of modifications to this model to improve and stabilize performance: we (i) extend the framework to curvilinear transducer arrays by accounting for element orientation~\cite{frey2025ultraflex}, (ii) utilize differentiable bent-ray refraction correction using Fermat's principle to enable accurate distributed sound speed estimation and aberration correction for transducer arrays with lenses, (iii) use an adaptive quality metric grid that is dynamically scaled based on the global beamforming sound speed, and (iv) modified the objective function to use a deadzone regularization approach that promotes more realistic sound speed distributions. We apply these changes to a large-scale in vivo evaluation of autofocusing on human liver images. 

\section{Beamforming Model Theory}
\label{sec:model}

\begin{figure*}[t]
    \centering
    \includegraphics[width=\textwidth,clip,keepaspectratio]{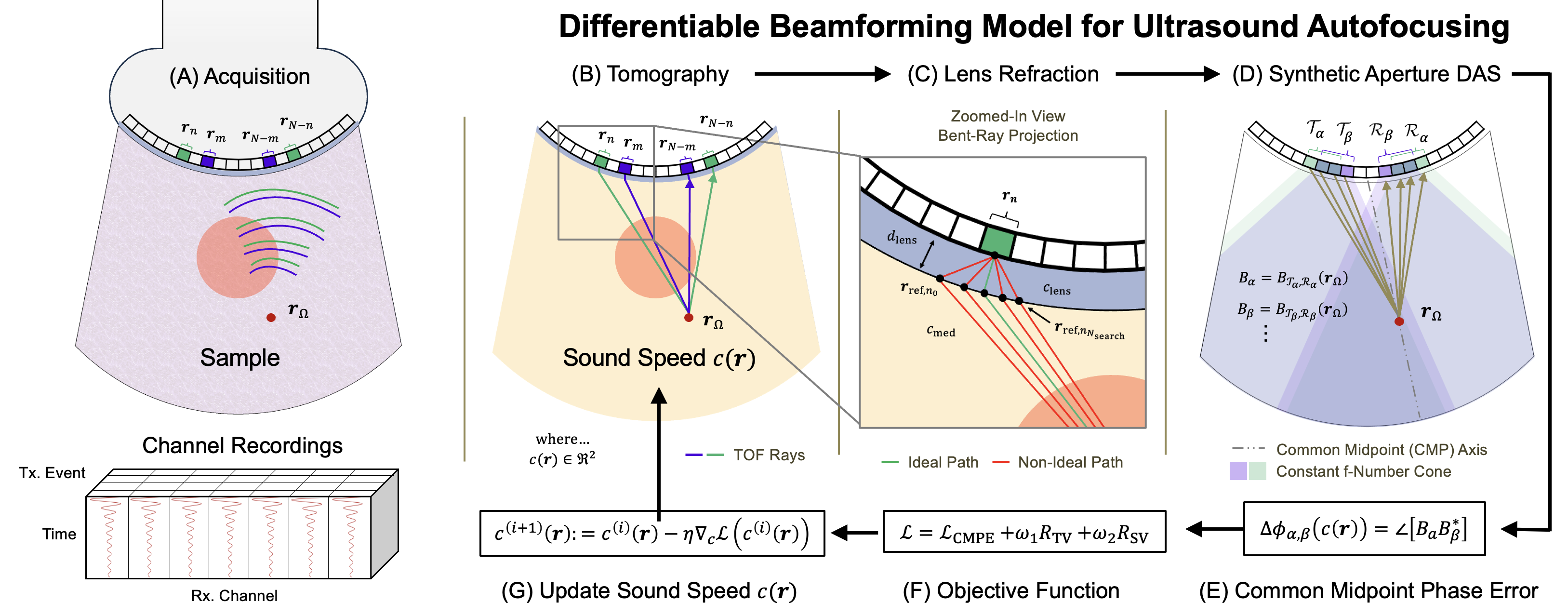}
    \caption{(a) Acquisition of the FSA data from a transducer array. The dark blue strip represents the lens of the transducer array. (b) Model for time-of-flight (TOF) calculation, where TOF values are determined through (c) bent-ray projection and refraction through the transducer lens using Fermat's principle. (d) Synthetic aperture DAS beamforming is performed using pairs of transmit and receive subapertures. (e) The common-midpoint phase-error (CMPE) focusing criterion is computed across beamformed subaperture images as part of the overall (f) objective function. (g) The objective function is differentiated with respect to the sound speed map, and the error is backpropagated through the beamforming model to update the distributed sound speed estimation.}
    \label{fig:diagram}
\end{figure*}

\subsection{Straight-Ray Model}
\label{sec:straight-ray}

The image reconstruction process utilizes full synthetic aperture (FSA) delay-and-sum (DAS) beamforming~\cite{jensen2006synthetic} to localize pulse-echo ultrasound signals spatially. The symbol $\mathbf{\Psi} \in \mathbb{C}^{N_s \times N_t \times N_r}$ will represent the demodulated FSA data. First (Fig.~\ref{fig:diagram}a), a time-of-flight (TOF) value from the $a$th element of a transducer array at position $\boldsymbol{r}_a$ (indexed by $n$ or $m$ over $N$ total elements in Fig.~\ref{fig:diagram}a) to a point in the target domain $\boldsymbol{r}_\Omega$ is calculated through a heterogeneous sound speed distribution $c(\boldsymbol{r})$:
\begin{equation}
    t_d(c(\boldsymbol{r}); \boldsymbol{r}_a, \boldsymbol{r}_\Omega) = \int_{\boldsymbol{r}_a}^{\boldsymbol{r}_\Omega} \frac{d\boldsymbol{r}}{c(\boldsymbol{r})},
    \label{eq:td}
\end{equation}
where $\boldsymbol{r}_a$ and $\boldsymbol{r}_\Omega$ are static variables (i.e., they are not updated over iterations). The round-trip TOF from a transmitting element position $\boldsymbol{r}_t$, to a point in the target domain $\boldsymbol{r}_\Omega$, and back to a receiving element $\boldsymbol{r}_r$ is $\tau_{t,r}(c(\boldsymbol{r}); \boldsymbol{r}_t, \boldsymbol{r}_r, \boldsymbol{r}_\Omega) = t_d(c(\boldsymbol{r}); \boldsymbol{r}_t, \boldsymbol{r}_\Omega) + t_d(c(\boldsymbol{r}); \boldsymbol{r}_\Omega, \boldsymbol{r}_r)$. For notational simplification, we will represent the combination of all transmit-receiver element pairs as $\vec{\boldsymbol{r}}_a$, so that all round-trip TOFs to a target domain position can be represented as $\tau=\tau_{t,r}(c(\boldsymbol{r}); \vec{\boldsymbol{r}}_a, \boldsymbol{r}_\Omega)$.

To accommodate a constant $f$-number for curvilinear arrays, we define spatial-weighting functions or spatial apodization for each element. We perform a transformation of $\boldsymbol{r}_\Omega$ into the element's frame of reference using a translation and rotation:
\begin{equation}
    \boldsymbol{r}_\Omega' = \begin{bmatrix}x_\Omega'\\ z_\Omega'\end{bmatrix} = 
    \begin{bmatrix}
    \cos\theta(\boldsymbol{r}_a) & -\sin\theta(\boldsymbol{r}_a) \\
    \sin\theta(\boldsymbol{r}_a) & \cos\theta(\boldsymbol{r}_a)
    \end{bmatrix}
    \begin{bmatrix}
    x_\Omega - x_a \\
    z_\Omega - z_a
    \end{bmatrix},
    \label{eq:rotation}
\end{equation}
where we infer the element orientation $\theta_a$ from neighboring elements positions as
\begin{equation}
    \theta(\boldsymbol{r}_a) = -\tan^{-1}\left((z_{a+1} - z_{a-1})/(x_{a+1} - x_{a-1})\right). 
    \label{eq:finite-difference}
\end{equation}
Finally, the spatial apodization of the $a$th array element using a constant $f$-number (denoted by $F$) is 
\begin{equation}
    h_a(\boldsymbol{r}_\Omega) = |x_\Omega'| \le \frac{F}{2} z_\Omega'   
    \label{eq:h},
\end{equation}
such that transmit and receive spatial apodizations are denoted  $h_t(\boldsymbol{r}_\Omega)$ and $h_r(\boldsymbol{r}_\Omega)$, respectively.

Incorporating Eqs.~\ref{eq:td}--\ref{eq:h} into the synthetic aperture DAS beamforming produces the straight-ray model of DAS beamforming through heterogeneous sound speed $c(\boldsymbol{r})$:
\begin{equation}
    B_{\mathcal{T},\mathcal{R}}(c(\boldsymbol{r}); \boldsymbol{r}_\Omega, \mathbf{\Psi}) = \sum_{t\in\mathcal{T}} \sum_{r\in\mathcal{R}} \mathbf{\Psi}_{t, r}(\tau) \exp(j2\pi f_d \tau) h_t h_r,
    \label{eq:beamforming}
\end{equation}
where $f_d$ is the demodulation frequency.

Phase rotation is performed by the exponential term in Eq.~\ref{eq:beamforming} to compensate for the demodulation-induced phase offset at the demodulation frequency. The beamformed signal $B$ can be formulated without a position $\boldsymbol{r}_{\Omega}$ in the target domain or the demodulated FSA data $\boldsymbol{\Psi}$ because these arguments are static across all model iterations. Notably, this beamforming implementation supports curvilinear arrays of any shape~\cite{frey2025ultraflex}. 

\subsection{Bent-Ray Model for Lens Compensation}

In Eq.~\ref{eq:beamforming}, it was assumed that the ray path was linear. This is a valid beamforming approximation when changes in sound speed are small ($\le$10\%) \cite{ali2022layered}. However, when a physical lens is introduced, this approximation no longer holds because of the large mismatch in sound speed between the lens and tissue. One approach is to compensate for lens effects by applying a manufacturer-provided scalar offset for lens correction $\hat\tau_{\text{lens cor}}$ (usually based on the 1540 m/s homogeneous sound speed assumption) and a scalar offset for the received waveform envelope $\hat\tau_{\text{ttp}}$ as
\begin{equation}
    \tau_{\text{Conventional}}=\tau_{t,r}(c(\boldsymbol{r}); \vec{\boldsymbol{r}}_a, \boldsymbol{r}_\Omega)+\hat\tau_{\text{lens cor}}+\hat\tau_{\text{ttp}}.
    \label{eq:conventional-tau}
\end{equation}
We will refer to this lens correction approach in Eq.~\ref{eq:conventional-tau} as the ``conventional" model. The advantage of this approach is that it is efficient to implement. However, this approach does not sufficiently model refraction through the lens or angular variations in lens thickness (i.e., an array element will see a greater effective lens thickness as the angle with respect to the element normal increases).

Another approach is to compensate for lens refraction by performing two-layer ray tracing based on Fermat's principle~\cite{waltham1988two}. Waasdorp et al.~\cite{waasdorp2024assessing} previously demonstrated that this approach enables accurate global sound speed estimation for simulated and physical transducers with varying lens thicknesses and sound speeds. In this case, illustrated by Fig.~\ref{fig:diagram}b, the lens is modeled as a layer of finite thickness $d_{\text{lens}}$ and sound speed $c_\text{lens}$, and the medium is modeled to have some global uniform sound speed $c_\text{med}$:
\begin{equation}
    \tau_{\text{BR}}=\tau_{\text{RayTracer}}(\hat{d}_{\text{lens}}, \hat{c}_{\text{lens}}, \hat{c}_{\text{med}})+\hat{\tau}_{\text{ttp}}.
    \label{eq:bent-ray-tau}
\end{equation}
We will refer to this ray-tracing approach as the ``bent-ray (BR)" model. We use these terms for ease of clarity in this paper because they can take on different meanings in the literature. The ray tracing term is effectively a bi-directional two-ray (four rays total) version of the straight-ray model that was introduced in Sec.~\ref{sec:straight-ray}. That is, Fermat's principle is used to find the path of shortest time from a transmitting element position $\boldsymbol{r}_t$, to a refractive position along the lens surface $\boldsymbol{r}_{\text{ref},t}$, to a target domain position $\boldsymbol{r}_\Omega$, to another refractive position along the lens surface $\boldsymbol{r}_{\text{ref},r}$, and finally to a receiving element position $\boldsymbol{r}_r$:
\begin{equation}
\begin{aligned}
    \tau_{\text{RayTracer}}:=t_d(\hat{c}_\text{lens}; \boldsymbol{r}_t, \boldsymbol{r}_{\text{ref},t})
    + t_d(c(\boldsymbol{r}); \boldsymbol{r}_{\text{ref},t}, \boldsymbol{r}_{\Omega})
    \\
    + t_d(c(\boldsymbol{r}); \boldsymbol{r}_{\Omega}, \boldsymbol{r}_{\text{ref},r})
    + t_d(\hat{c}_\text{lens}; \boldsymbol{r}_{\text{ref},r}, \boldsymbol{r}_{r}).
    \label{eq:ray-tracer}
\end{aligned}
\end{equation}

\subsection{Sound Speed Estimation}

Modern machine learning approaches that utilize neural networks are based on automatic differentiation frameworks. These frameworks are powerful because they enable gradient-based parameter optimization of many distinct physics-based models. Our approach is similar to the training of a neural network; we optimize an underlying physical parameter of our beamforming model (i.e., distributed sound speed) instead of the weights and biases of a neural network. In this approach, we utilize common-midpoint phase error as a focusing criterion~\cite{simson2025ultrasound}, and we backpropagate loss through our differentiable image reconstruction process to update our original estimate of distributed sound speed.

Our model for distributed aberration correction (Fig.~\ref{fig:diagram}) starts with an FSA transmit sequence shown in Fig.~\ref{fig:diagram}a. The model takes in demodulated FSA data and performs bent-ray tracing (Figs.~\ref{fig:diagram}b-c) through the lens of the transducer based on global domain sound speed. The global sound speed used for model initialization is determined by assessing the common-midpoint phase error (CMPE) through a discrete range of global sound speed values. The sound speed with the lowest resulting CMPE is used as a global initialization value for the autofocusing model. 

In Simson et al.~\cite{simson2025ultrasound} and this work, the sound speed distribution is represented by a discrete grid of sound speed value ``pixels". However, in Simson et al., a fixed assumption was made about grid pixel spacings using a static sound speed value of 1500 m/s. Because a range of sound speed values are swept during model initialization, the pixel locations and kernel spacings effectively evaluate different time samples, thereby measuring CMPE from different speckle signals for each sound speed. In this work, we have replaced this fixed (static) assumption with a dynamic approach that updates the CMPE pixel and kernel samples based on the active sound speed value used in the sweep. With this dynamic grid approach, the CMPE kernels observe the same speckle at each global beamforming sound speed.

Next, the autofocusing model is initialized with this optimized global sound speed value represented by an initially uniform discrete grid. Straight rays are again projected from the lens surface, and sound speed values from the discrete grid are interpolated onto ray segments for TOF calculation. These TOF values are used to perform synthetic aperture DAS beamforming (Fig.~\ref{fig:diagram}), and then the CMPE is used as a focusing criterion in the objective function (Figs.~\ref{fig:diagram}e-f)~\cite{simson2025ultrasound}. Finally, the partial derivative of the objective function with respect to the sound speed distribution is calculated and backpropagated through the entire differentiable beamforming framework to update the original distributed sound speed estimation. The update step for iteration $i$ is given as:
\begin{equation}
    c(\boldsymbol{r})^{(i+1)} = c(\boldsymbol{r})^{(i)} 
    - \gamma \frac{\partial \mathcal{L}}{\partial c(\boldsymbol{r})} 
    \left(c(\boldsymbol{r})^{(i)} \right),
    \label{eq:update-rule-component}
\end{equation}
where $\gamma$ is the step size and $\partial \mathcal{L}(c(\boldsymbol{r})^{(i)}) / \partial c(\boldsymbol{r})$ denotes the gradient of the objective function $\mathcal{L}$ with respect to the sound speed distribution evaluated at iteration $i$. The sound speed distribution is updated in the negative gradient direction in this case because we seek to minimize the objective function. This process is repeated iteratively for a pre-defined number of iterations $N_i$. While we use a fixed number of iterations in our evaluations here, a convergence criterion could be utilized in the future to stop the optimization process when the updates no longer produce significant changes.

\subsection{Objective Function}

The common-midpoint phase error (CMPE) metric was introduced as a focusing criterion by Simson et al.~\cite{simson2025ultrasound} for ultrasound autofocusing via sound speed estimation. Here, we retain the same CMPE implementation as the focusing loss $\mathcal{L}_{\mathrm{CMPE}}$ and regularize the estimated sound speed distribution using thresholded first- and second-order spatial variation. Defining the deadzone function as $D_{\kappa}(x)=\max(0,|x|-\kappa)$, the first-order regularization is

\begin{equation}
    R_{\mathrm{TV}}(c)
    =
    \frac{1}{N}\sum_{i,j}
    \left[
    D_{\kappa_x^{(1)}}\left(\partial_x c_{i,j}\right)
    +
    D_{\kappa_z^{(1)}}\left(\partial_z c_{i,j}\right)
    \right].
    \label{eq:tv_reg}
\end{equation}

To suppress high-frequency spatial oscillations, we similarly define the second-order regularization as

\begin{equation}
    R_{\mathrm{SV}}(c)
    =
    \frac{1}{N}\sum_{i,j}
    \left[
    D_{\kappa_x^{(2)}}\left(\partial_{xx} c_{i,j}\right)
    +
    D_{\kappa_z^{(2)}}\left(\partial_{zz} c_{i,j}\right)
    \right].
    \label{eq:sv_reg}
\end{equation}

Here, $\partial_x$, $\partial_z$, $\partial_{xx}$, and $\partial_{zz}$ denote the corresponding first- and second-order finite-difference operators in physical coordinates. The thresholds $\kappa_x^{(1)}$ and $\kappa_z^{(1)}$ specify allowable sound speed gradients, while $\kappa_x^{(2)}$ and $\kappa_z^{(2)}$ specify allowable curvatures. Thus, only spatial variations exceeding the prescribed thresholds are penalized. Both terms are evaluated in physical units and averaged over the spatial domain. The complete optimization objective is

\begin{equation}
    \mathcal{L}
    =
    \mathcal{L}_{\mathrm{CMPE}}
    +
    w_1 R_{\mathrm{TV}}(c)
    +
    w_2 R_{\mathrm{SV}}(c),
    \label{eq:objective}
\end{equation}
where $w_1$ and $w_2$ control the first- and second-order
regularization strengths, respectively.

\section{Methods}
\label{sec:methods}

\subsection{Model Implementation}

The ultrasound autofocusing process via sound speed estimation is shown in Fig.~\ref{fig:diagram}. A discrete ``slowness" grid is used to represent the inverse of the continuous sound speed distribution (i.e., $\boldsymbol{s}\approx1/c(\boldsymbol{r})$). In the straight-ray model, rays are projected from the element positions and continue to target-domain positions $\boldsymbol{r}_\Omega$. However, in the bent-ray model, rays are projected from element positions, refract through the lens, and then continue to target-domain positions. After leaving the lens, the straight rays are composed of $N$ segments with $N+1$ segment endpoints. Slowness values are interpolated onto the segment endpoint positions using bi-cubic interpolation. Finally, a discrete summation is used to calculate TOF values through the heterogeneous sound speed distribution.

In Fig.~\ref{fig:diagram}b, there exists ${\left( N_\text{search} \right)^2}$ combinations of bent-ray paths from a transmitting element position $\boldsymbol{r}_t$, to a target domain position $\boldsymbol{r}_\Omega$, to a receiving element position $\boldsymbol{r}_r$. During model initialization, we assume the sound speed in the medium to be homogeneous, such that $c_{\text{med}}=\overline{c(\boldsymbol{r})}$, where the overline represents the mean sound speed value in the medium. From these ${\left( N_\text{search} \right)^2}$ path combinations, the path of minimum TOF is selected as the ideal transmit and receive refractive positions, $\boldsymbol{r}_{\text{ref},t}$ and $\boldsymbol{r}_{\text{ref},r}$, respectively. Although the local sound speed distribution is iteratively updated through the differentiable beamforming model, we determine these refractive positions for each iteration using the current global-average sound speed of the medium. 

Before the iterative process is started, the medium is initialized as a homogeneous sound speed distribution. This sound speed value is determined by performing a sweep of the CMPE across a range of sound speed values from 1340 to 1740 m/s. A dynamic grid of pixel locations is used for CMPE calculation based on the active sound speed value, with each pixel location containing a 3 x 3 kernel with $\lambda/2$ spacing to mitigate the effect of jitter on the CMPE calculation. The autofocusing model was implemented in JAX~\cite{jax2018github}.

\subsection{Simulations}

Six two-dimensional targets with varying sound speed and density distributions were produced using k-Wave~\cite{treeby2010k} and QUPS~\cite{brevett2024qups}: two uniform, two circular inclusion, and two ``TMI" logo sound speed layouts. All simulations used an FSA sequence with a transducer based on the Verasonics $\mbox{C5-2v}$ curvilinear transducer array (Verasonics Inc., Kirkland, WA, USA) with the following parameters: $N_a=128$ elements, 49.6 mm radius of curvature, $0.5872^\circ$ angular pitch ($508$ $\mu$m linear pitch), center frequency $f_c=3.7$ MHz, and 70$\%$ bandwidth. The simulation medium was discretized using a $\lambda/8$ grid spacing. The transducer lens was also simulated in each case. The simulations were performed using the k-Wave GPU binaries on an NVIDIA RTX A6000 (NVIDIA, Santa Clara, CA) with a sample rate of 105 MHz. The resulting RF channel data were demodulated, low-pass filtered, and downsampled by a factor of 24 for the uniform targets and 22 for inclusion/"TMI" targets. The demodulated FSA data was passed into the model.

The six targets comprised three iso-impedance and three anisotropic-impedance simulations. This paired design was intended to assess whether the sound speed estimation component of the aberration correction model would be significantly degraded in the presence of echogenicity variations. Iso-impedance refers to impedance matching at the lens-background and feature-background, but not the scatterers. The anisotropic-impedance simulations were created using different density and sound speed distributions (i.e., an image of a keyboard was used for the density map). For the uniform layout, the sound speed is 1650 m/s, and the nominal density is 884.8 kg/m$^3$ for the iso-impedance version, and 1000.0 kg/m$^3$ for the keyboard-impedance version. For the inclusion and ``TMI" layouts, the background sound speed is 1460 m/s, and the nominal background density is 1000 kg/m$^3$. The feature (i.e., either the inclusion or ``TMI" letters) sound speed is 1500 m/s, and nominal feature density is 973.3 kg/m$^3$ for the iso-impedance versions, and 1000.0 kg/m$^3$ for the keyboard-impedance versions. In the case of the ``TMI" logo in the sound speed map and the keyboard image in the density map, the natural image is loaded into $\text{MATLAB}$, converted to grayscale, and thresholded to create a binary mask. Random amplitude variations from a uniform distribution were applied to the density map to emulate sub-wavelength scatterers, assuming a scatterer density of 30 reflectors per $\lambda^2$.

All six targets included a curvilinear lens. The simulated transducer elements were placed behind the lens arc at a uniform distance of 0.90 mm to emulate a finite lens thickness. The lens thickness and sound speed were modeled after calibration values acquired from an experimental $\mbox{C5-2v}$ array (see Sec.~\ref{sec:calibration}). In all cases, the lens sound speed is 999.4 m/s, and the density is 1460.8 kg/m$^3$. Additionally, between the lens and the medium, a gradient of thickness $\lambda/4$ was implemented to minimize numerical instability; the lens thickness was correspondingly reduced (i.e., because the lens has a lower sound speed than the medium) to preserve the overall acoustic travel time. 

\subsection{Phantom and in vivo Human Liver Data}

RF channel data were previously acquired from four sound speed phantoms using a Verasonics (Verasonics Inc., Kirkland, WA, USA) \mbox{L12-3v} linear transducer array and Verasonics Vantage 256 research ultrasound system. The \mbox{L12-3v} array has 192 elements, 0.2 mm pitch, and a nominal center frequency of 7.8 MHz. FSA data were previously acquired with a 6.0 MHz transmit center frequency from three graphite-gelatin phantoms (phantoms 1--3 have calibrated sound speeds 1490 m/s, 1516 m/s, and 1540 m/s, respectively)~\cite{ali2022layered}. A chicken breast (calibrated sound speed of 1575 m/s) was placed on top of each phantom to produce a bi-layer phantom. FSA data (only the center 128 elements were captured) were also previously acquired with a 7.35 MHz transmit center frequency from a CIRS 040GSE phantom (manufacturer-provided sound speed of 1540 m/s)~\cite{frey2024sgc}.

Additional RF channel data were acquired from two phantoms for this work using Verasonics \mbox{L12-3v} linear and \mbox{C5-2v} curvilinear transducer arrays on a Verasonics Vantage 256 system. The two phantoms were the CIRS 040GSE phantom and a CIRS 049A elastography phantom (1546 m/s calibrated sound speed from a manufacturer sample). RF channel data were acquired using the \mbox{L12-3v} array (CIRS 049A only; only 128 elements were captured) with a 7.35 MHz transmit frequency and an FSA sequence. RF channel data were also acquired using a \mbox{C5-2v} array (128 elements, 49.6 mm radius of curvature, 0.51 mm pitch) using a 4.0 MHz transmit frequency and an FSA sequence from both phantoms. 

Eighty-one high-BMI subjects from Stanford Hospital \& Clinics were enrolled in this study under a Stanford IRB approved protocol (protocol number IRB-56630). After informed consent was obtained, in vivo human liver data was collected using the \mbox{C5-2v} transducer with the same settings as previously described for the phantoms. Of the 81 subjects, data was only collected from 76 subjects (comprising 321 acquisitions). Eight improper acquisitions were excluded, leaving 313 acquisitions for analysis. All phantom and in vivo data were demodulated and low-pass filtered before being passed into the ``bent-ray'' model.

\subsection{Transducer Lens Calibration}
\label{sec:calibration}

We used the transducer lens parameter estimation process previously described in Waasdorp et al.~\cite{waasdorp2024assessing}. Briefly, this lens parameter calibration procedure is performed by executing a highly repetitive FSA sequence (200 frames in our case) while the transducer is air-coupled. For the case of the simulated C5-2v array, we used a single-frame FSA sequence with a grid spacing of $\lambda/8$ to model the internal reflection caused by the lens/air interface. In post-processing, we then average transmit events to form a parabolic time trace as a function of element position (i.e., the transmitting elements are placed at the origin). Finally, a multi-parametric optimization of $c_{\text{lens}}$, $d_{\text{lens}}$, and $t_{\text{ttp}}$ is performed by maximizing the coherence factor of the channel data from a parabolic curve fit to the time-delay values as in~\cite{waasdorp2024assessing}.

\begin{table}[ht]
    \centering
    \renewcommand{\arraystretch}{1}
    \caption{Summary of Lens Calibration Measurements}
    \begin{tabular}{lccc}
        \toprule
        \textbf{Transducer Array} & \textbf{$c_{\text{lens}}$ (m/s)} & \textbf{$d_{\text{lens}}$ (mm)} & \textbf{$t_{\text{ttp}}$ (ns)} \\
        \midrule
        L12-3v   & 1049.4 & 0.67 & 578.3 \\
        C5-2v (Phantoms)  & 999.4  & 0.90 & 1328.6 \\
        C5-2v (Humans)   & 990.9  & 0.83 & 1210.2 \\
        \bottomrule
    \end{tabular}
    \label{tab:calibration}
\end{table} 

In this paper, one \mbox{L12-3v}, one simulated \mbox{C5-2v}, and two \mbox{C5-2v} arrays were used. Their measured calibration values are provided in Table~\ref{tab:calibration}. Of the two C5-2v arrays, one was used for laboratory phantom acquisitions, and the other for in vivo subject acquisitions. All transducers were calibrated using the same Vantage 256 system. When the autofocusing model is run on the simulations, the parameters $c_{\text{lens}}$ = 999.4~m/s and $d_{\text{lens}}$ = 0.90~mm corresponding to the \mbox{C5-2v} ``phantoms" array was used. In the k-Wave simulations, the waveform peak is nearly aligned with the negative start time of the simulation, such that $t_{\text{ttp}}$ = 0.

\subsection{Autofocusing Model Hyperparameters}

\begin{table}[t]
  \centering
  \caption{Hyperparameters used for simulation,\\phantom, and in vivo experiments}
  \setlength{\tabcolsep}{4pt}
  \begin{tabular}{lcccc}
    \toprule
    Parameter & Simulation$^*$ & \multicolumn{2}{c}{Phantom} & In Vivo \\
    \cmidrule(lr){3-4}
     & & L12-3v & C5-2v & C5-2v \\
    \midrule
    LR                         & 0.0023  & 0.0030 & 0.0030 & 0.0500 \\
    $w_1$                      & 13.38    & 7.00     & 10.00    & 100.00 \\
    \quad$\kappa_x^{(1)}$     & 0.4225  & 0.0010 & 0.0100  & 0.0100 \\
    \quad$\kappa_z^{(1)}$     & 0.0009 & 0.02000 & 0.0100  & 0.0100 \\
    $w_2$                      & 3.31     & 0     & 100.00   & 100.00 \\
    \quad$\kappa_x^{(2)}$     & 0.6615  & 0     & 0.0010 & 0.0020 \\
    \quad$\kappa_z^{(2)}$     & 0.3523  & 0     & 0.0010 & 0.0020 \\
    $\rho_{\text{thresh}}$    & 0.9506  & 0.9000   & 0.9000   & 0.9000 \\
    $F$                        & 0.1683  & 0.7500  & 0.5000   & 0.4000 \\
    $N_{\text{segments}}$     & 368      & 32    & 32    & 64 \\
    \bottomrule
    \multicolumn{5}{p{0.95\columnwidth}}{\footnotesize
    $^*$ These values have been rounded to the nearest significant digit.} \\
  \end{tabular}
  \label{tab:hyperparameters}
\end{table}

In Table~\ref{tab:hyperparameters}, the learning rate (LR) controls the optimization step size; $w_1$ and $w_2$ weight the total-variation ($R_{\text{TV}}$) and second-order variation ($R_{\text{SV}}$) regularizers, respectively. The parameters $\kappa_x^{(1)}$ and $\kappa_z^{(1)}$ limit the lateral and axial first derivatives of sound speed, whereas $\kappa_x^{(2)}$ and $\kappa_z^{(2)}$ limit the corresponding second derivatives. The remaining parameters are the correlation-based filtering threshold $\rho_{\text{thresh}}$, optimization $f$-number $F$, and number of ray segments $N_{\text{segments}}$. For the set of hyperparameters used for experiments with the simulated data, low weighting was assigned to the second-order variation regularization term $w_2$, a low $f$-Number $F$, and a large number of ray segments $N_{\text{segments}}$ were used because the inclusion and "TMI" were intricate geometries with large sound speed discontinuities. This was also the case for the layered phantoms under the \mbox{L12-3v} column, which used no second-order variation regularization.

\subsection{Model Evaluation}
The mean absolute error (MAE) in sound speed was used to assess sound speed estimations for simulation and calibrated phantom experiments:
\begin{equation}
    \text{MAE} = \frac{1}{N_\Omega} \sum_{i=1}^{N_\Omega} |c_{\text{est}, v} - c_{\text{true}, v}|
\end{equation}
where $c_{\text{est}}$ and $c_{\text{true}}$ are the model-estimated and ground truth sound speeds, respectively.  

Contrast, contrast-to-noise ratio (CNR), speckle brightness~\cite{nock1989phase}, coherence factor~\cite{mallart1994adaptive}, lag-one coherence~\cite{long2018lag}, common-midpoint correlation coefficient, and common-midpoint phase error~\cite{simson2025ultrasound} were used to assess aberration correction for in vivo human liver experiments. Each of these quality metrics are described in more detail in Frey et al.~\cite{frey2025ultraflex}. Contrast, in units of decibels (dB), and CNR were calculated as described by Rodriguez-Molares et al.~\cite{rodriguezmolares2020generalized}.

For each image, one foreground ROI was selected using a polygon selection tool, and four background patches outside (above, below, and on both sides) the ROI were selected using a rectangular selection tool. This process was repeated for both the corrected and aberrated image version. Then, for calculation of the contrast and CNR values, the average of the four contrast (or CNR) measurements between the foreground and background was calculated for both the corrected and aberrated image separately. This approach was used because targets changed positions after autofocusing due to variations in sound speed used for reconstruction. The global image quality metrics were computed as described in Frey et al. \cite{frey2025ultraflex}.

The Tenengrad sharpness metric \cite{vraalstad2026coherence} was used to assess the relative change in image sharpness before and after aberration correction and is calculated in two steps. First, a lateral Sobel filter is applied to the log-compressed B-mode images. For this metric calculation, each corrected log-compressed B-mode image is normalized so that its 80th-percentile pixel amplitude corresponds to the 80th-percentile pixel amplitude in the aberrated log-compressed B-mode. This was to maintain consistency with Vr{\aa}lstad et al.~\cite{vraalstad2026coherence}. Next, the sum $S$ of all pixel amplitudes in the filtered image is calculated beginning at a depth of 10 mm. Finally, the relative change $\eta$ between the corrected image sharpness $S_{\text{corrected}}$ and aberrated image sharpness $S_{\text{aberrated}}$ is calculated as $\eta=100\left(S_{\text{corrected}}/S_{\text{aberrated}}-1\right)$.

\section{Results}
\label{sec:results}

\subsection{Simulation}

\begin{figure}[ht]
    \centering
    \includegraphics[width=\columnwidth,clip,keepaspectratio]{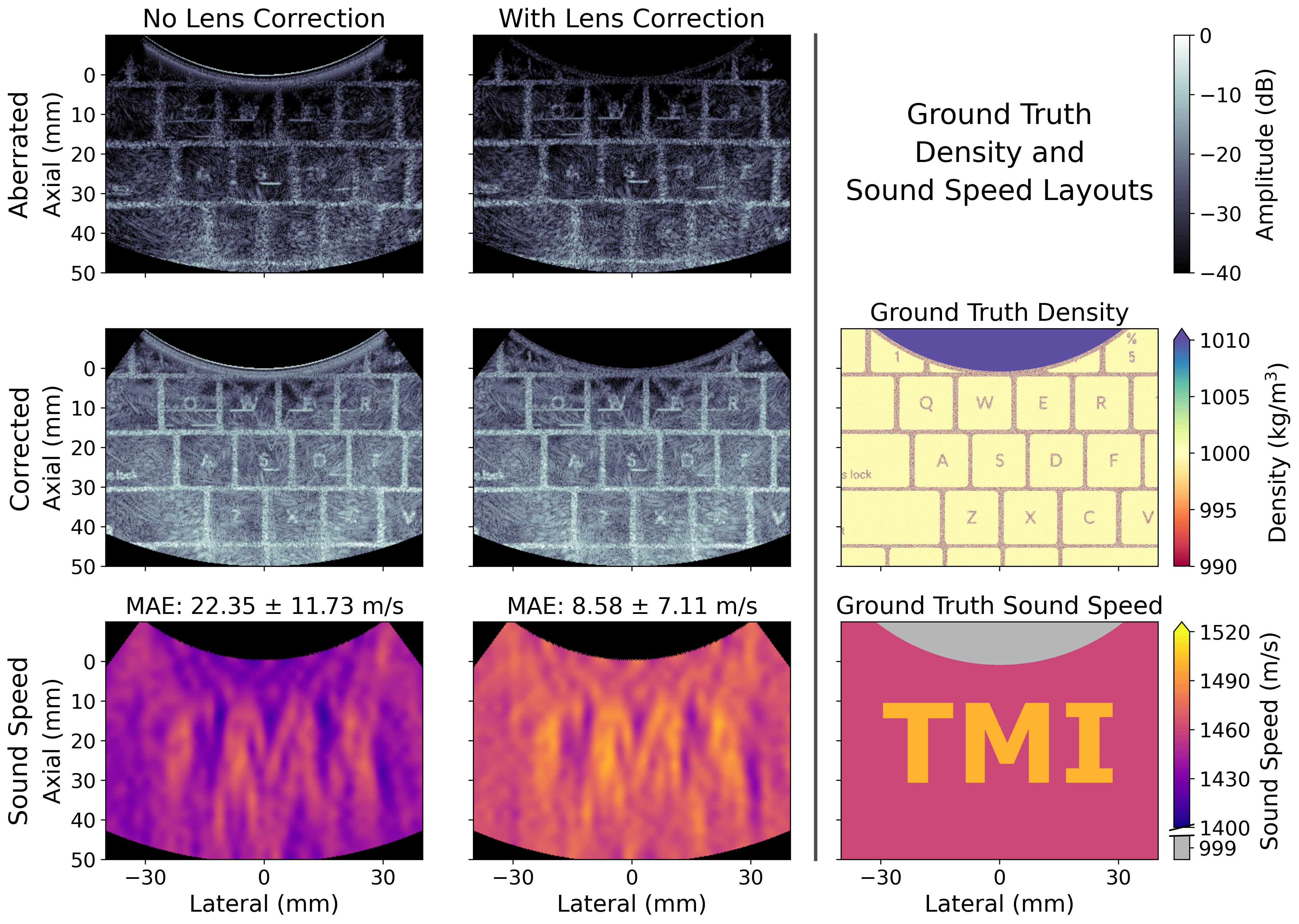}
    \caption{Model results on simulation data for the "TMI" logo target with a keyboard-based impedance map. The top row depicts the B-mode reconstructed with a 1540 m/s homogeneous sound speed, the middle row shows B-mode reconstruction using the distributed sound speed estimation produced by the autofocusing model, and the bottom row shows the estimated sound speed distribution. Each column pair compares results (left) without lens correction with (right) the bent-ray lens correction model. The last column shows the ground truth density and sound speed maps for reference.}
    \label{fig:simulation-results}
\end{figure}

Fig.~\ref{fig:simulation-results} compares aberration correction and sound speed estimation for the autofocusing model with and without lens correction for the "TMI" logo target. Aberration correction significantly improved image brightness. The bent-ray lens correction process reduced beamforming artifacts in the B-mode near the transducer surface. Furthermore, the bent-ray lens correction produced more accurate sound speed estimates (especially near the transducer arc) and lower MAE sound speed values for all simulation targets. MAE in the estimated sound speed for each target was 6.3 m/s (25.3 m/s) on average for the uniform targets, 7.3 m/s (21.0 m/s) on average for the inclusion targets, and 7.8 m/s (23.4 m/s) on average for the “TMI” logo targets with (without) lens correction.

\subsection{Phantom}

\begin{table}[t]
  \centering
  \caption{Phantom mean absolute errors (MAEs) in m/s}
  \label{tab:phantom_mae_simplified}
  \begin{tabular}{llcc}
    \toprule
    Phantom & Probe & Simson et al. & Ours \\
    \midrule
    Phantom 1 & L12-3v & 27.3$\pm$17.0 & 26.2$\pm$21.4 \\
    Phantom 2 & L12-3v & 19.1$\pm$9.0 & 16.6$\pm$10.8 \\
    Phantom 3 & L12-3v & 9.9$\pm$4.8 & 9.7$\pm$6.0 \\
    CIRS 040GSE & L12-3v & 17.2$\pm$13.0 & 8.1$\pm$3.1 \\
     & C5-2v & 29.1$\pm$10.7$^\dagger$ & 11.4$\pm$3.3 \\
    CIRS 049A & L12-3v & 14.0$\pm$12.1 & 13.0$\pm$3.5 \\
     & C5-2v & 38.7$\pm$12.2$^\dagger$ & 2.2$\pm$1.8 \\
    \bottomrule
    \multicolumn{4}{p{0.95\columnwidth}}{\footnotesize
    For phantoms 1-3, the reported MAE is across the full field of view (both the chicken and the gelatin phantom). $^\dagger$Represents our model without lens correction; the Simson et al. model does not account for curvilinear arrays.} \\
  \end{tabular}
\end{table}

Aberration correction models and sound speed estimates were validated on data acquired from calibrated tissue-mimicking phantoms and compared against the previous version of the autofocusing model~\cite{simson2025ultrasound}. In Table~\ref{tab:phantom_mae_simplified}, the bent-ray model enabled a reduction in average MAE from 17.5 m/s to 14.7 m/s for phantom data acquired with the linear array (\mbox{L12-3v}). For data acquired with the curvilinear array (\mbox{C5-2v}), the model reduced average MAE from 33.9 m/s to 6.8 m/s.

\subsection{In Vivo Human Liver}

\subsubsection{Model Initialization}

\begin{figure}[ht]
    \centering
    \includegraphics[width=\columnwidth,clip,keepaspectratio]{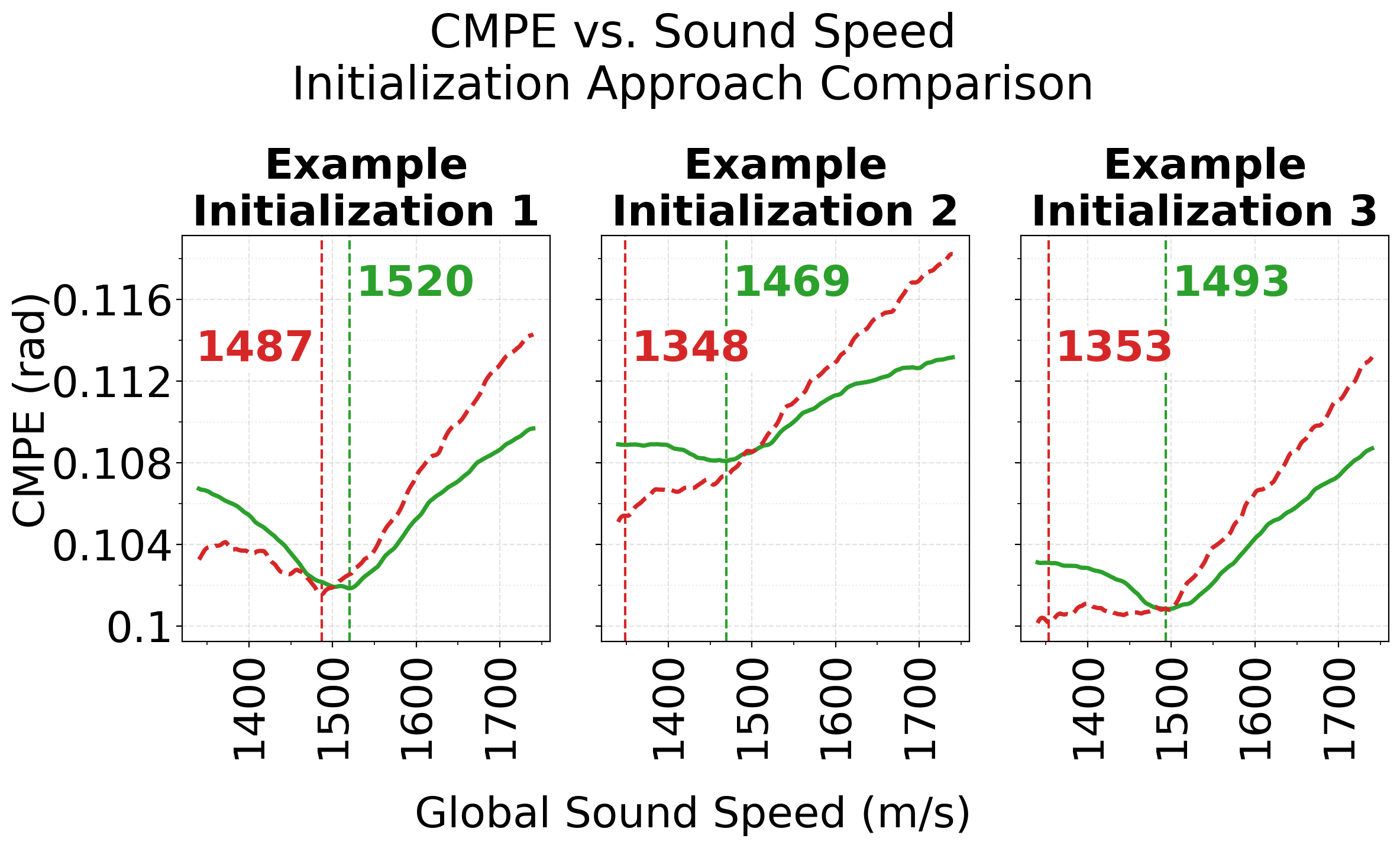}
    \caption{Comparison of the static grid (red curve) and dynamic grid (green curve) approaches used for model initialization. The vertical lines represent the selected optimal global sound speed.}
    \label{fig:init}
\end{figure}

In Fig.~\ref{fig:init}, a comparison of the previous static and updated dynamic approaches is shown for three example in vivo cases. This model modification had a significant effect on the convex-optimal sound speed value used for global sound speed initialization. Most notably, the static approach was susceptible to poor model initialization and sometimes lacked a convex profile resulting in selection of the lowest sound speed value in the sweep. The dynamic approach fixes this issue by sampling the same temporal samples through dynamic adjustment of CMPE pixel and kernel positions based on sound speed used for beamforming. 

\begin{figure*}[t]
    \centering
    \includegraphics[width=\textwidth,clip,keepaspectratio]{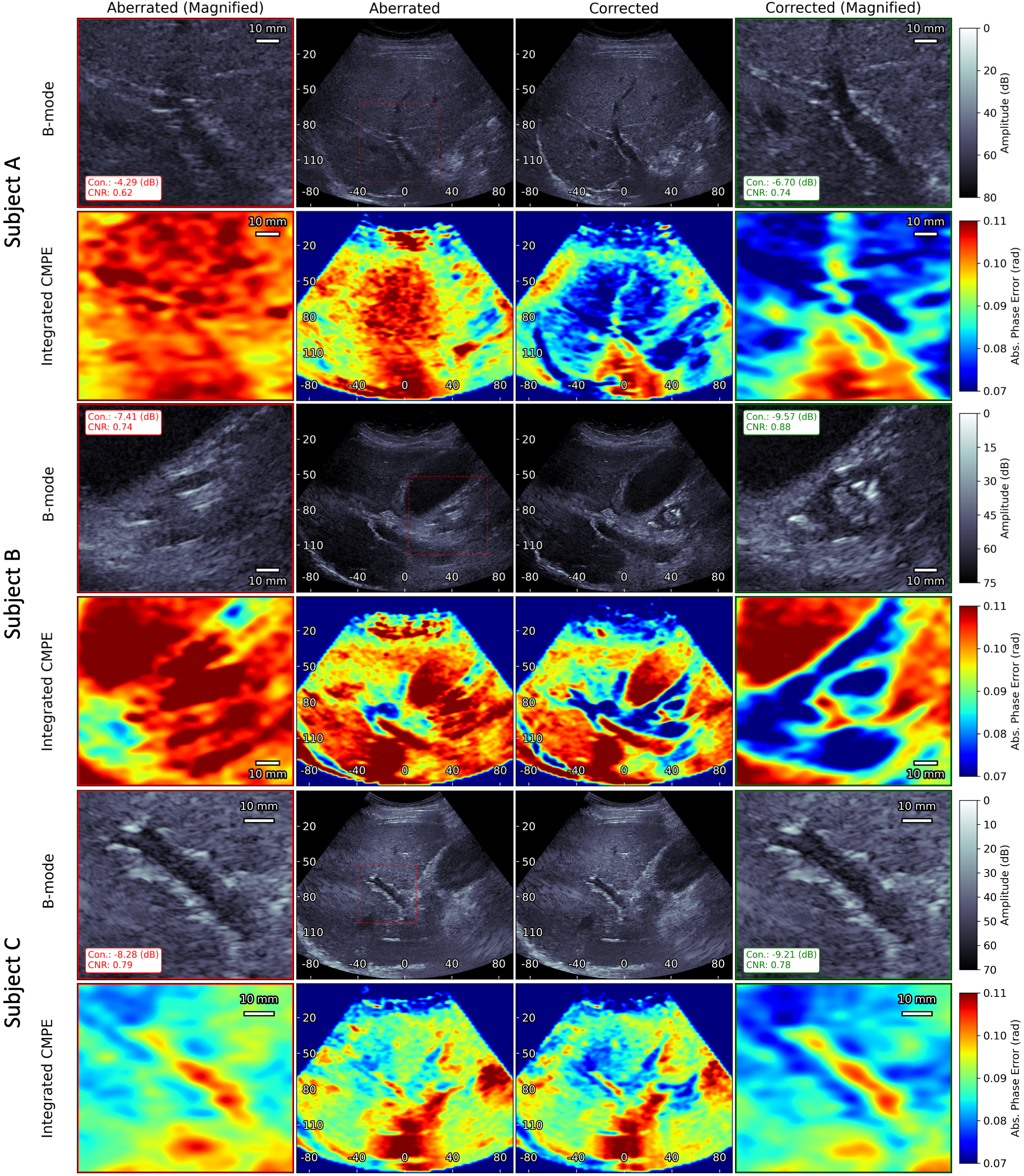}
    \caption{Model results on in vivo human liver data from three different subjects. Rows 1, 3, and 5 depict B-mode reconstruction with a 1540 m/s homogeneous sound speed on the left (aberrated) and the B-mode reconstruction using the updated autofocusing model on the right (corrected). Rows 2, 4, and 6 depict the integrated CMPE as an image corresponding to the B-mode region shown above it.}
    \label{fig:liver-results}
\end{figure*}

\subsubsection{Autofocusing Results}

The qualitative results of aberration correction on three in vivo human liver cases are shown in Fig.~\ref{fig:liver-results}. In Fig.~\ref{fig:liver-results}, Subjects A and B show significant aberration correction in the B-mode image. Below each B-mode image is a corresponding integrated CMPE image, which is the phase-sum of all CMPE produced by all transmit/receive subaperture pairs at each pixel location. The left columns of Fig.~\ref{fig:liver-results} show the aberrated images (using the conventional model for reconstruction with a 1540 m/s homogeneous sound speed assumption), while the right columns show the corrected image reconstructions using the updated autofocusing model. The inner columns show the full curvilinear image field-of-view, whereas the outer columns show an inset of the image feature in the red box. For subject A, the walls of the main portal vein (red box) increase in visibility and definition. For this subject, there is a corresponding overall significant reduction in CMPE throughout the entire image, especially in the tissue surrounding the portal vein and from 10 to 20 mm depth. For subject B, the gallbladder becomes darker after aberration correction, and the shape of the small intestine below the gallbladder is revealed (red box). The corresponding integrated CMPE images show a significant reduction of the phase error in tissue surrounding the gallbladder and small intestine. Subject C is a representative case in which the autofocusing model does not improve B-mode image quality. The phase error across the entire domain was already relatively low before aberration correction, and despite some phase error reduction around the hypoechoic region on the right side of the image, there is no change in the quality of the reconstruction of the B-mode. The walls of the anechoic feature (pictured in the inset) did not change in width. 

Contrast and CNR improvement are observed in the highlighted targets in two of the three cases. For subject A, contrast and CNR improved from -4.29 dB to -6.70 dB and 0.62 to 0.74, respectively. For subject B, contrast and CNR improved from -7.41 dB to -9.57 dB and 0.74 to 0.88, respectively, in the small intestine (positioned below the gallbladder). For subject C, contrast improved from -8.28 dB to -9.21 dB, whereas CNR decreased slightly from 0.79 to 0.78. The MAE in the final estimated sound speed from 1540 m/s was 56.9 m/s for Subject A, 30.0 m/s for Subject B, and only 9.4 m/s for Subject C. Although we have used the term MAE here, these values should not be viewed as an error, but as a deviation from the assumption of 1540 m/s.

\begin{figure*}
    \centering
    \includegraphics[width=\textwidth,clip,keepaspectratio]{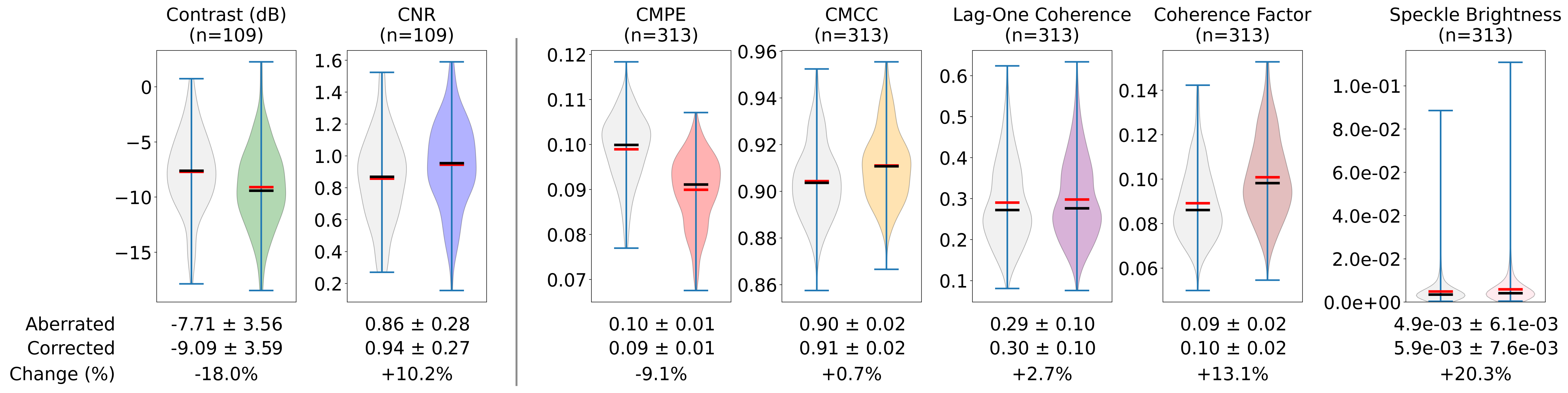}
    \caption{Quantitative measurements across all in vivo human liver data involving contrast, CNR, common midpoint phase error (CMPE), common midpoint correlation coefficient (CMCC), lag-one coherence, coherence factor, and speckle brightness. In each case, the gray boxes represent metric results from data that was beamformed at 1540 m/s (i.e., no aberration or lens correction), whereas the colored boxes represent metric results from data that was beamformed using the bent-ray autofocusing model. The red line represents the mean and the black line represents the median. The mean and standard deviation of each violin plot is provided below the plot. }
    \label{fig:metrics}
\end{figure*}

The quantitative results of aberration correction on the in vivo data reveal average contrast and CNR changes of -1.38$\pm$1.60 dB (-18.0\%, lower is better) and 0.09$\pm$0.14 (+10.2\%), respectively, across 109 annotated targets in Fig.~\ref{fig:metrics}. Across all 313 acquisitions, the mean values of the global image-quality metrics improved. The global image quality metrics included speckle brightness (+20.3\%), coherence factor (+13.1\%), lag-one coherence (+2.7\%), common-midpoint correlation coefficient (+0.7\%), and common-midpoint phase error (-9.1\%, lower is better). The paired differences departed from normality for several metrics. Therefore, a two-sided paired Wilcoxon signed-rank test was performed comparing aberrated and corrected reconstructions consistently across all metrics. The changes in contrast, CNR, and all global image-quality metrics were statistically significant ($p<10^{-6}$).

\begin{figure}[ht]
    \centering
    \includegraphics[width=\columnwidth,clip,keepaspectratio]{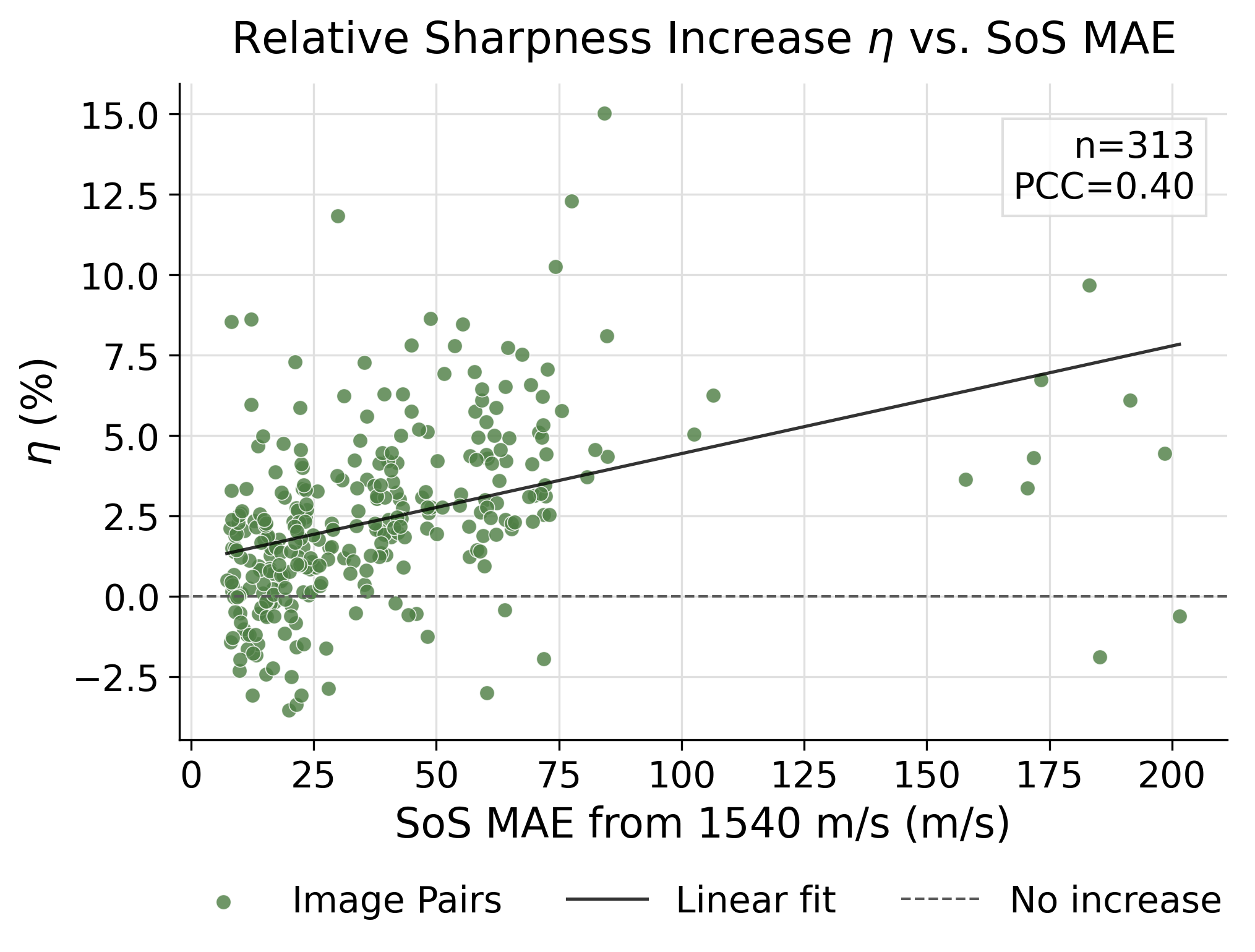}
    \caption{Relative increase in the Tenengrad sharpness metric for aberrated and corrected image pairs ($n=313$).}
    \label{fig:tenengrad}
\end{figure}

The mean Tenengrad sharpness increased by $2.4\%$ after aberration correction, and this change was statistically significant according to a two-sided paired Wilcoxon signed-rank test ($p<10^{-6}$). Fig.~\ref{fig:tenengrad} reveals a positive Pearson correlation coefficient ($\rho$=0.40, $n=313$, $p<10^{-6}$) between the relative change in sharpness $\eta$ and the MAE in the final estimated sound speed from 1540 m/s. The aberration-correction process generally results in a positive change in sharpness (i.e., a relative change above $\eta=0\%$), although 52 cases resulted in a negative change in sharpness. Four cases showed a greater than 10\% increase in relative sharpness, while 11 cases had a MAE in estimated sound speed from 1540 m/s greater than 100 m/s.

\section{Discussion}
\label{sec:discussion}

In the simulation data, the bent-ray lens correction model enabled the lowest sound speed estimation MAE across all six targets. The qualitative sound speed distribution also improved; without lens correction, a large sound speed gradient extending from a depth of 0 to 20 mm is apparent. This artifact is eliminated in the cases where the bent-ray lens correction was used. This near-field error is important because inaccurate sound speed estimates close to the transducer can influence aberration throughout the rest of the image. Our bent-ray lens correction mitigates this issue by more accurately accounting for (compared to a pure straight-ray assumption) propagation through the lens and the shallow region beneath the array lens. 

For the anisotropic targets, distributed sound speed estimation with high resolution is challenging due to weak common midpoint signals received from the anechoic keys while using a fixed CMPE evaluation grid. This limitation is apparent in regions where the method has weak signal to optimize (i.e., a CMPE evaluation patch was positioned in a hypoechoic key), resulting in a less-reliable sound speed estimation. In Table~\ref{tab:hyperparameters}, the set of parameters used for optimization is reported for each dataset. However, parameters can be fine-tuned for each target to improve the result, but such an approach would be impractical for real-time imaging or clinical settings.

Regarding aberration correction, for the "TMI" logo without lens correction, there is refraction at the lens-tissue interface, which produces angle-dependent delay errors that appear as a near-field halo. This artifact is eliminated in the bent-ray model. Second, even without lens correction, the ability to resolve the individual keyboard letters is significantly improved through aberration correction. For example, in the anisotropic "TMI" logo example, in the bottom keyboard row, only the "V" key is legible. However, after aberration correction, all keys in the same row are legible. Note that the "TMI" logo example represents beamforming with a speed above that of the domain, resulting in axial stretching of the keys in the anisotropic-impedance case. After aberration correction is applied, the keys are the correct height. 

Our findings are similar for sound speed estimation ability on experimental phantom data (Table~\ref{tab:phantom_mae_simplified}). While linear arrays can benefit from lens correction for more accurate sound speed estimation, curvilinear array with lens correction undergo even more drastic sound speed estimation correction due to the more severe angles at which the ray paths intersect the lens.

Regarding aberration correction results for experimental in vivo human liver data, we demonstrate substantial aberration correction across different views of the liver. Aberration correction can be observed both qualitatively in the presented B-modes and integrated CMPE images, as well as quantitatively via an improvement in contrast, CNR, all global image quality metrics, and Tenengrad sharpness. However, the visual improvement in image quality was often more closely tied to reductions in CMPE. To some extent, this is expected because our autofocusing method directly minimizes CMPE. However, the images showing large correction also tended to have significant CMPE reduction close to the transducer surface (e.g. \ref{fig:liver-results}, Subjects A \& B), whereas ones that did not have significant CMPE reduction in this region tended not to have significant image distortion (e.g. \ref{fig:liver-results}, Subject C). In regions containing speckle-generating tissue, the autofocusing model reduces CMPE, indicating improved spatial coherence and better alignment of the received echoes after correction. This reduction is consistent with the observed sharpening of specular reflectors and improved target visibility. However, CMPE does not necessarily improve in anechoic or low signal regions such as the gallbladder in the top right of Subject B, Fig.\ref{fig:liver-results}, or the lung in the bottom left corner. In these regions, the received signals contain limited backscattered energy, so the CMPE calculation is dominated by noise or lack of signal. Therefore, autofocusing improvements are most meaningful in tissue regions that provide sufficient signal, while anechoic regions may remain unchanged. 

Visual improvement in image quality was also closely tied to increasing MAE in the estimated sound speed from 1540 m/s. For example, Subjects A and B exhibited significant aberration correction and had final estimated sound-speed distributions with large MAE values of 56.9 m/s and 33.0 m/s from 1540 m/s, respectively. Whereas, Subject C showed no visible aberration correction and had a smaller MAE value of 9.4 m/s from 1540 m/s.

Finally, traditional contrast and CNR metrics are useful for assessing target intensity with respect to background intensity, but they do not adequately quantify improvements in geometric distortion or target definition. Therefore, these evaluation metrics are limited in their ability to assess aberration correction methods, especially for small targets or when high contrast targets are not present.

\section{Conclusion}
\label{sec:conclusion}

Image distortion in clinical ultrasound scanner systems is primarily caused by wavefront aberration. This work addresses two technical challenges facing the practical implementation of clinical aberration correction by estimating the underlying local tissue sound speed distribution and correcting for lens refraction using a differentiable bent-ray model. We also proposed a new initialization approach for the model based on an adaptive grid to account for the change in speckle position as a function of the sound speed. These changes enabled the first large-scale in vivo validation of our distributed aberration-correction method on 313 liver acquisitions from 76 high-BMI human subjects. For subject images with anechoic regions, contrast and CNR improved on average. Multiple image quality metrics were evaluated: speckle brightness, coherence factor, lag-one coherence, common-midpoint correlation coefficient, and common-midpoint phase error. Through our aberration correction process, all quality metrics improvements were statistically significant. These results demonstrate the potential for future clinical distributed aberration correction techniques using ultrasound autofocusing.

\section*{Acknowledgment}
We would like to thank the Maresca and Renaud labs and Dr. Rick Waasdorp for providing Verasonics acquisition scripts and discussion related to his lens calibration work.

\bibliographystyle{IEEEtran}

\bibliography{abbrev,refs}

\end{document}